\documentclass[12pt]{article}
  \usepackage[english]{babel}
\usepackage[pdftex]{color}
\usepackage{amsmath}
\usepackage{graphicx}
\usepackage{hyperref}
\usepackage{amsmath}
\usepackage{amsfonts}
\usepackage{amssymb}
\usepackage{url}
\newcommand {\pom} {I\!\! P}
\begin{document}

\title{
{\bf{} MBR Monte Carlo Simulation in PYTHIA8}}

\author{Robert Ciesielski, Konstantin Goulianos\\
  {\small The Rockefeller University, 1230 York Avenue, New York, NY 10065, USA} 
   \\
  {\small E-mail: robert.ciesielski@rockefeller.edu, dino@rockefeller.edu}
}

\date{}

\maketitle

\begin{abstract}
We present the {\bf\sc mbr} (Minimum Bias Rockefeller) Monte Carlo simulation of (anti) proton-proton interactions and its implementation in the {\sc pythia8} event generator. We discuss the total, elastic, and total-inelastic cross sections, and three contributions from diffraction dissociation processes that contribute to the latter: single diffraction, double diffraction, and central diffraction or double-Pomeron exchange. The event generation follows a renormalized-Regge-theory model, successfully tested using CDF data. Based on the {\bf\sc mbr}-enhanced {\sc pythia8} simulation, we present cross-section predictions for the {\bf\sc lhc} and beyond, up to collision energies of 50 TeV.  
\end{abstract}

\section{Introduction}
\begin{figure*}[!tb]
  \begin{center}
    \includegraphics[width=0.8\textwidth]{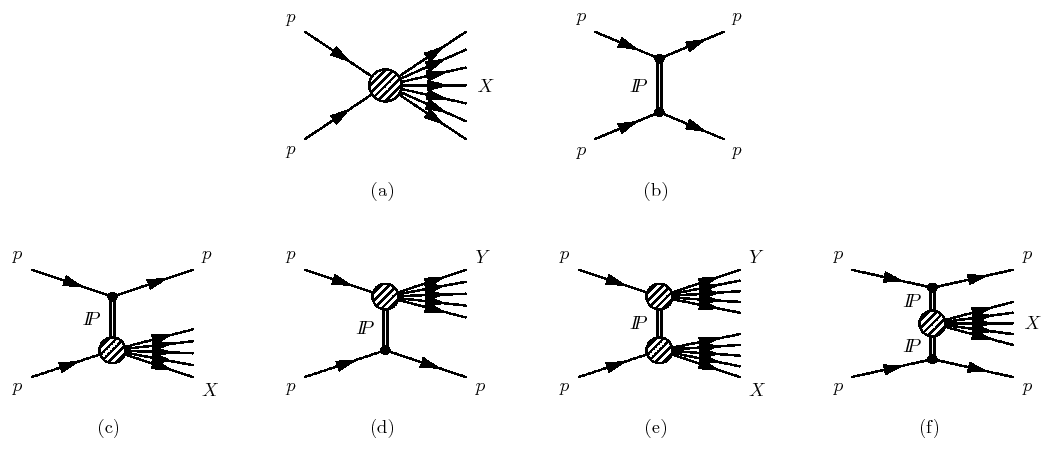}
    \caption{Schematic diagrams of soft $pp$ processes addressed in {\bf\sc mbr}: (a) total cross section, (b) elastic scattering, (c)-(d) single diffraction or single dissociation (SD), (e) double diffraction or double dissociation (DD), and (f) central diffraction (CD) or double-Pomeron exchange (DPE).}
    \label{fig:1}
  \end{center}
\vspace*{-1em}\end{figure*}

The {\bf\sc mbr} (Minimum Bias Rockefeller) Monte Carlo (MC) simulation is an event generator addressing the contributions of three diffraction-dissociation processes to the total-inelastic $pp$ cross section\footnote{Although the original {\bf\sc mbr} addresses several hadron-hadron collisions, including $\bar pp$, we will be assuming $pp$ collisions throughout this paper for simplicity, except as explicitly stated.}: single-diffraction dissociation or single dissociation (SD), in which one of the incoming protons dissociates, double-diffraction dissociation or double dissociation (DD), in which both protons dissociate, and central dissociation (CD) or double-Pomeron exchange (DPE), where neither proton dissociates. These processes are tabulated below,
\begin{eqnarray}
{\rm SD}& pp\rightarrow Xp\nonumber\\
{\rm or}&pp\rightarrow pY\nonumber\\
{\rm DD}&pp\rightarrow XY\nonumber\\
{\rm CD}\;{\rm (DPE)}&pp\rightarrow pXp,\nonumber
\label{eqn:processes}
\end{eqnarray}
where $X$ and $Y$ represent diffractively dissociated protons. Schematic diagrams are shown in Fig.~\ref{fig:1}, along with diagrams for the total and elastic cross sections, which are also simulated in {\bf\sc mbr}. 

{\bf\sc mbr} predicts the energy dependence of the total, elastic, and total-inelastic $pp$ cross sections, and fully simulates the above three diffractive components of the total-inelastic cross section. The diffractive-event generation is based on a phenomenological renormalized-Regge-theory model\cite{dinoModel}, originally developed for the CDF experiment. We have implemented {\bf\sc mbr} in {\sc pythia8}\cite{p8ref}, where it can be activated with the flag: \texttt{Diffraction:PomFlux} = 5.

This paper is organized as follows: in Sec.~\ref{xsecs} we present the formulae used to calculate the total and elastic cross sections, as well as the diffractive and non-diffractive components of the total-inelastic cross section; in Sec.~\ref{evgen} we describe details of the generation of diffractive events; and in Sec.~\ref{cards} we summarize the {\sc pythia8} ``steering cards'' for employing {\bf\sc mbr} in the simulation.

\section{Cross sections}
\label{xsecs}

The total cross section ($\sigma_{\rm tot}$) is the sum of the total-elastic ($\sigma_{\rm el}$) and total-inelastic ($\sigma_{\rm inel}$) cross sections. The $\sigma_{\rm tot}$ and $\sigma_{\rm el}$ are calculated in Sec.~\ref{TotElXsec}. The total-inelastic cross section ($\sigma_{\rm inel}$) is obtained as $\sigma_{\rm inel} = \sigma_{\rm tot} - \sigma_{\rm el}$. The $\sigma_{\rm inel}$ receives contributions from the diffractive components (SD, DD, CD or DPE) and from the non-diffractive (ND) cross section, defined as:
\begin{equation}
\sigma_{\rm ND}=(\sigma_{\rm tot} - \sigma_{\rm el}) - (2\sigma_{\rm SD}+\sigma_{\rm DD}+\sigma_{\rm CD}),
\label{eqND}
\end{equation}
where $\sigma_{\rm SD}$ is the cross section for either $pp\rightarrow Xp$ or $pp\rightarrow pY$, assumed to be equal. The diffractive cross sections are calculated in Sec.~\ref{diffXsec}.

\subsection{Total, elastic, and total-inelastic cross sections}
\label{TotElXsec}

The $\sigma^{p^{\pm}p}_{\rm tot}(s)$ cross sections at a $pp$ center-of-mass-energy $\sqrt s$ are calculated as follows:
\begin{equation}
\sigma^{p^{\pm}p}_{\rm tot} =
\begin{cases}
 16.79 s^{0.104} + 60.81 s^{-0.32} \mp 31.68 s^{-0.54}& \text{for } \sqrt{s} < 1.8 \mbox{ TeV},\\
 \sigma_{\rm tot}^{\rm CDF}+\frac{\pi}{s_0}\left[ \left(\ln\frac{s}{s_F}\right)^2- \left(\ln\frac{s^{\rm CDF}}{s_F} \right)^2\right] & \text{for } \sqrt{s} \ge 1.8 \mbox{ TeV},
\end{cases}
\label{eqTOT}
\end{equation}  

The term for $\sqrt s<1.8$~TeV, where the $(\pm)$ denotes $\left(^p_{\bar p}\right)$,  is obtained from a global Regge-theory fit to pre-LHC data 
on $p^{\pm}p,\,K^{\pm }p$ and $\pi^{\pm }p$ cross sections~\cite{globalFit}, while that for $\sqrt s\ge 1.8$~TeV is a prediction of a model based on a saturated Froissart bound\cite{satFroiss}. The latter, which  we normalize to the CDF measurement of $\sigma_{\rm tot}$ at $\sqrt{s^{\rm CDF}}=1.8$ TeV,  $\sigma_{\rm tot}^{\rm CDF}=80.03 \pm 2.24~\mbox{mb}$,
depends on two parameters: the energy at which the saturation occurs, $\sqrt{s_F}= 22~\mbox{GeV}$, and the energy-scale parameter, $s_0$, for which we use $s_0=(3.7 \pm 1.5)\,{\rm GeV}^2$, divided by $(\hbar c)^2\approx 0.389\;{\rm GeV}^2{\rm mb}$ to obtain the cross section in Eq.~(\ref{eqTOT}) in mb.

The elastic cross section, $\sigma^{p^{\pm}p}_{\rm el}$, is calculated using $\sigma_{\rm tot}$ from Eq.~(\ref{eqTOT}), multiplied by the elastic-to-total cross-section ratio, $\sigma_{\rm el}/\sigma_{\rm tot}$, obtained from the global Regge fit of Ref.~\cite{globalFit}. The total inelastic cross section is calculated as $\sigma_{\rm inel}=\sigma_{\rm tot}-\sigma_{\rm el}$. 

\begin{figure*}[t]
  \begin{center}
    \includegraphics[width=0.9\textwidth]{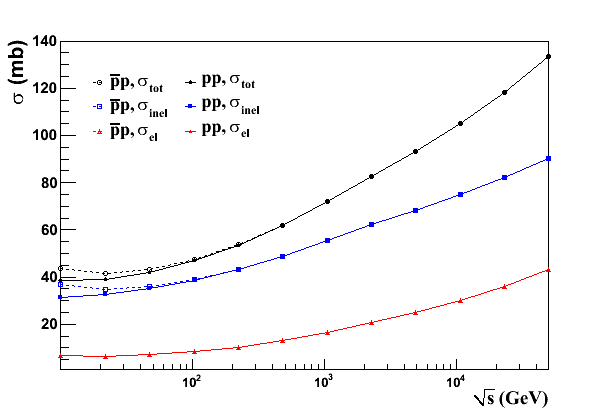}
    \caption{Total, elastic and total-inelastic $pp$ and $\bar{p}p$ cross sections vs. $\sqrt{s}$.}
    \label{fig:2}
  \end{center}
\end{figure*}

The energy dependences of $\sigma_{\rm tot}$, $\sigma_{\rm el}$ and $\sigma_{\rm inel}$ are shown in Fig.~\ref{fig:2}, and cross-section values at $\sqrt{s}$=0.3, 0.9, 1.96, 2.76, 7, 8 and 14 TeV are presented in Tab.~\ref{tab:1}.

\subsection{Diffractive cross sections}
\label{diffXsec}

Cross sections for SD, DD and CD (or DPE) are calculated using a phenomenological model discussed in detail in Ref. \cite{dinoModel}. Differential cross sections are expressed in terms of the Pomeron ($\pom$) trajectory,  $\alpha(t)=1+\epsilon +\alpha't = 1.104 + 0.25~(\mbox{GeV}^{-2})\cdot t$, the Pomeron-proton coupling, $\beta(t)$, and the ratio of the triple-$\pom$ to the $\pom$-proton couplings, $\kappa \equiv g(t)/\beta(0)$. For sufficiently large rapidity gaps ($\Delta y \gtrsim $ 3), for which $\pom$-exchange dominates, the cross sections may be written as,  
\begin{eqnarray}
\frac{d^2\sigma_{SD}}{dtd\Delta y} & = & \frac{1}{N_{\rm gap}(s)} \left[ \frac{~ ~ \beta^2(t)}{16\pi}e^{2[\alpha(t)-1]\Delta y}\right] \cdot \left\{ \kappa \beta^2(0) \left( \frac{s'}{s_{0}}\right)^{\epsilon}\right\}, \label{eqSD}\\
\frac{d^3\sigma_{DD}}{dtd\Delta y dy_0} & = & \frac{1}{N_{\rm gap}(s)} \left[ \frac{\kappa\beta^2(0)}{16\pi}e^{2[\alpha(t)-1]\Delta y}\right] \cdot \left\{ \kappa \beta^2(0) \left( \frac{s'}{s_{0}}\right)^{\epsilon}\right\}, \label{eqDD}\\
\frac{d^4\sigma_{DPE}}{dt_1dt_2d\Delta y dy_c} & = & \frac{1}{N_{\rm gap}(s)} \left[\Pi_i\left[ \frac{\beta^2(t_i)}{16\pi}e^{2[\alpha(t_i)-1]\Delta y_i}\right]\right] \cdot \kappa \left\{ \kappa \beta^2(0) \left( \frac{s'}{s_{0}}\right)^{\epsilon}\right\}, \label{eqCD}
\end{eqnarray}
where $t$ is the square of the four-momentum-transfer at the proton vertex and $\Delta y$ is the rapidity gap width. The variable $y_0$ in Eq.~(\ref{eqDD}) is the center of the rapidity gap. In Eq.~(\ref{eqCD}), the subscript $i=1, 2$ enumerates Pomerons in the DPE event, $\Delta y=\Delta y_1 + \Delta y_2$ is the total (sum of two gaps) rapidity-gap width in the event, and $y_c$ is the center in $\eta$ of the centrally-produced hadronic system. 

Eqs.~(\ref{eqSD}) and (\ref{eqDD}) are equivalent to those of standard-Regge theory, as $\xi$, the fractional forward-momentum-loss of the surviving proton (forward momentum carried by $\pom$), is related to the rapidity gap by $\xi=e^{-\Delta y}$. 
The variable $\xi$ is defined as $\xi_{\rm SD}=M^2/s$ and $\xi_{\rm DD}=M_1^2M_2^2/(s\cdot s_0)$, where $M^2$ ($M_1^2$, $M_2^2$) are the masses of dissociated systems in SD (DD) events. For DD events, $y_0=\frac{1}{2}\ln(M_2^2/M_1^2)$, and for  DPE $\xi=\xi_1\xi_2=M^2/s$.

The Pomeron-proton coupling , $\beta(t)$, is given by:
\begin{equation}
\beta^2(t)=\beta^2(0)F^2(t),
\end{equation}
where $\beta(0)= 4.0728 ~ \sqrt{\mbox{mb}} = 6.566 ~ \mbox{GeV}^{-1}$ and $F(t)$ is the proton form factor from Ref. \cite{donnachie}:
\begin{equation}
F^2(t)=\left[ \frac{4m^2_p-2.8t}{4m^2_p-t}\left(\frac{1}{1-\frac{t}{0.71}} \right)^2\right]^2 \approx a_1e^{b_1t}+a_2e^{b_2t}.
\label{eqFF}
\end{equation}
The right-hand side of Eq.~(\ref{eqFF}) is a double-exponential approximation of $F^2(t)$, with $a_1=$0.9, $a_2$=0.1, $b_1=$4.6 GeV$^{-2}$, and $b_2=$0.6 GeV$^{-2}$.
 The term in curly brackets in Eqs.(\ref{eqSD})-(\ref{eqCD}) 
is the $\pom$-$p$ total cross section at the reduced $\pom$-$p$ collision energy squared, $s'=s \cdot e^{-\Delta y}$. 
The parameter $\kappa$ is set to $\kappa=0.17$~\cite{kappa},  and $\kappa \beta^2(0)\equiv \sigma_0$, where $\sigma_o$ defines the total Pomeron-proton cross section at an energy-squared value  of $s_0 = 1~\mbox{GeV}^2$, is set to $\sigma_0= 2.82~\mbox{mb}$ or $7.249~\mbox{GeV}^{-2}$.

\begin{figure*}[b]
  \begin{center}
    \includegraphics[width=0.9\textwidth]{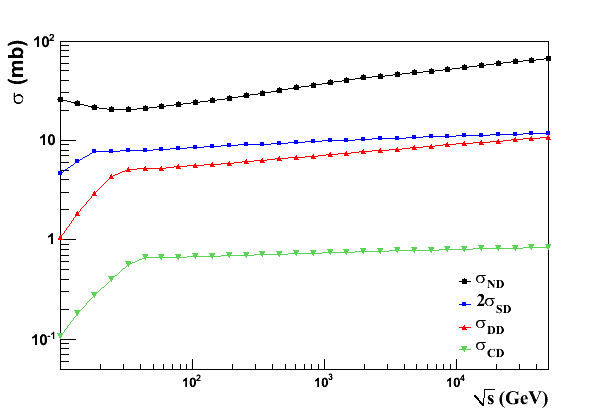}
    \caption{Diffractive (SD, DD, CD) and non-diffractive (ND) cross sections vs. $\sqrt{s}$.}
    \label{fig:3}
  \end{center}
\end{figure*}

\clearpage

Pointing again to Eqs.(\ref{eqSD})-(\ref{eqCD}, the renormalization parameter, $N_{\rm gap}(s)$, is defined as $N_{\rm gap}(s)=min(1,f)$, where $f$ is the integral of the term in square brackets, corresponding to a Pomeron flux. The integral is calculated over all phase space in $t_i$ and in $\eta_0$ (DD) or $\eta_c$ (DPE) for $\Delta y > 2.3$. The renormalization procedure we use, which is based on interpreting the Pomeron flux as a (diffractive) gap-formation probability, yields predictions which are in very good agreement with diffractive measurements at CDF\cite{cdfSD,cdfDD,cdfCD}.

The cross-section formulae are used to generate events with large (diffractive) rapidity gaps, $\Delta y$. We suppress cross sections at small value of $\Delta y$ by multiplying Eqs.~(\ref{eqSD})-(\ref{eqCD}) by:
\begin{equation}
S=\frac{1}{2}\left[ 1+erf\left(\frac{\Delta y - \Delta y_{S}}{\sigma_{S}}\right)\right],
\label{erffun}
\end{equation}
where $erf$ is the error function \cite{erfFun}, centered by default at $\Delta y_{S}=2$ with a width of $\sigma_{S}=0.5$. For SD, this choice corresponds to suppressing events above the coherence limit ($\xi  \lesssim 0.135$) \cite{coherLimit}. For DD, the choice of $\Delta y_{S}=2$ is somewhat arbitrary, because of the difficulty of unambiguously distinguishing a low-$\Delta y$ DD event from a ND event with a rapidity gap from (exponentially suppressed) fluctuations. Changing $\Delta y_{S}$ introduces event migrations between DD and ND samples. Such migrations  have no impact on the total-inelastic cross section, as can be inferred from Eq.~(\ref{eqND}). For DPE, the suppression is applied to the total-gap width, $\Delta y=\Delta y_1 + \Delta y_2$.

Fig.~\ref{fig:3} presents the energy dependence of diffractive cross sections calculated as explained above. The non-diffractive cross section, given by Eq.~(\ref{eqND}) is also shown. Cross-section values for $\sqrt{s}$=0.3, 0.9, 1.96, 2.76, 7, 8 and 14 TeV are tabulated in Tab.~\ref{tab:1}. 


\begin{table}[t!]
\begin{center}
\begin{tabular}{|c|c|c|c|c|c|c|c|}
\hline
$\sqrt{s}$ (TeV) &0.3&0.9& 1.96 & 2.76 & 7 & 8 & 14\\
\hline
$\sigma_{\rm tot}$ & 56.50 & 69.87 & 81.03 & 85.25 & 98.29 & 100.35 & 109.49\\
$\sigma_{\rm el}$ & 11.28 & 15.83 & 19.97 & 21.70 & 27.20 & 28.09& 32.10\\
$\sigma_{\rm inel}$ & 45.23 & 54.04 & 61.06 & 63.55 & 71.10 & 72.26 & 77.39\\
\hline
$\sigma_{ND}$ & 29.19 & 36.50 & 42.41 & 44.39 & 50.57 & 51.54 & 55.84\\
$\sigma_{2SD}$ & 9.10 & 9.76 & 10.22 & 10.41 & 10.91& 10.98& 11.26\\
$\sigma_{DD}$ & 6.21 & 7.03 & 7.67 & 7.97 & 8.82 & 8.94 & 9.47\\
$\sigma_{CD}$ & 0.718 & 0.746 & 0.766 & 0.776 & 0.800 & 0.804 & 0.818\\
\hline

\end{tabular}
\end{center}
\caption{Cross sections in {\em mb} of the processes contributing to the Minimum-Bias sample, calculated as explained in Sec.~\ref{TotElXsec} and Sec.~\ref{diffXsec},  for selected values of $\sqrt{s}$.}
\label{tab:1}
\end{table}

\section{Event generation}
\label{evgen}

\begin{figure*}[t!]
  \begin{center}
    \includegraphics[width=1.\textwidth]{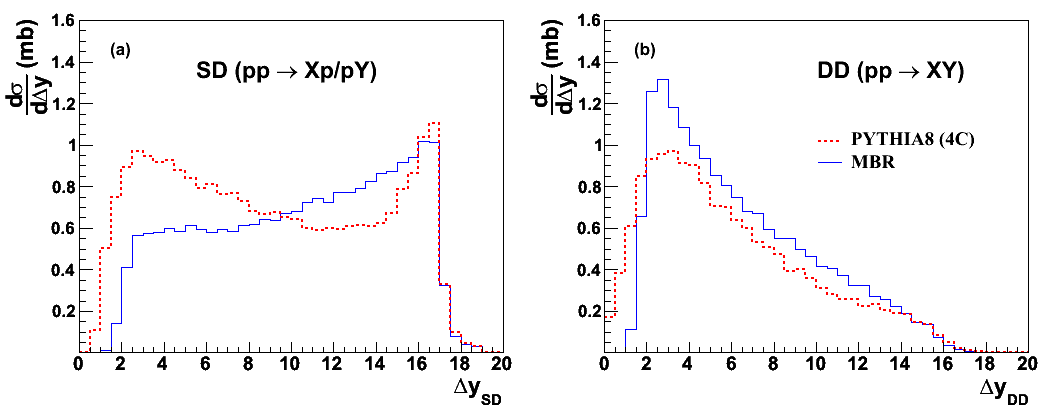}
    \caption{Differential cross sections as a function of $\Delta y$ for (a) SD and (b) DD at $\sqrt{s}$ = 7 TeV, compared to predictions of {\sc pythia8-4C} (rescaled Sch\"{u}ler \& Sj\"{o}strand model\cite{p8ref}, \texttt{Diffraction:PomFlux=1}).}
    \label{fig:4}
  \end{center}
\end{figure*}

\begin{figure*}[h!]
  \begin{center}
    \includegraphics[width=1.\textwidth]{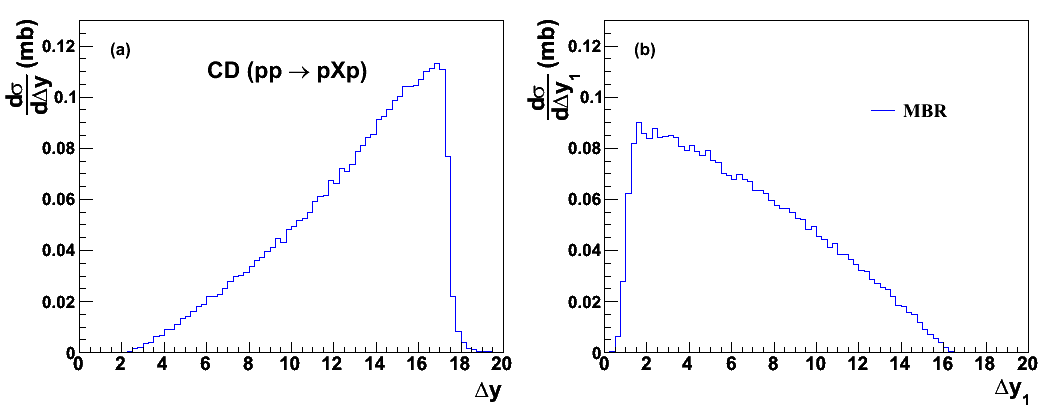}
    \caption{Differential cross sections for CD (DPE) at $\sqrt{s}$ = 7 TeV as a function of: (a) total-gap width ($\Delta y = \Delta y_1 + \Delta y_2$), and (b) single-gap width ($\Delta y_1$).}
    \label{fig:5}
  \end{center}
\end{figure*}

The four-momenta of particles produced in the event are generated as described below, in Secs.~\ref{genSD}-\ref{genCD}. The results of the simulation for SD and DD at $\sqrt{s}$ = 7 TeV are presented in Fig.~\ref{fig:4}, which shows differential cross sections as a function of rapidity-gap width, $\Delta y$, compared to predictions of {\sc pythia8-4c} (rescaled Sch\"{u}ler \& Sj\"{o}strand model\cite{p8ref}, \texttt{Diffraction:PomFlux = 1}). The distribution of $d\sigma/d\Delta y$ vs. the total rapidity-gap-width, $\Delta y = \Delta y_1 + \Delta y_2$, and vs. the width of an individual gap, $\Delta y_1$, for CD at $\sqrt{s}$ = 7 TeV is shown in Fig.~\ref{fig:5}.

\subsection{Single-diffractive events}
\label{genSD}

Events are generated by first choosing the rapidity-gap width, $\Delta y$, according to Eq.~(\ref{eqSD}) integrated over $t$:
\begin{equation}
\frac{d\sigma_{SD}}{d\Delta y} \sim e^{\epsilon \Delta y} \cdot \left( \frac{a_1}{b_1+2\alpha' \Delta y} + \frac{a_2}{b_2+2\alpha' \Delta y} \right) \cdot S.
\end{equation}
The range of the generation is defined by $\Delta y_{min}=0$ and $\Delta y_{max}=-\ln{M^2_0/s}$, where $M^2_0=$\texttt{{\bf\sc mbr}m2Min}. The term:
\begin{equation}
S=\frac{1}{2}\left[ 1+erf\left(\frac{\Delta y - \texttt{{\bf\sc mbr}dyminSD}}{\texttt{{\bf\sc mbr}dyminSigSD}}\right)\right],
\end{equation}
is added to suppress events at low values of $\Delta y$, as explained in Sec.~\ref{diffXsec}.

A value of $t$ is then chosen according to:
\begin{equation}
\frac{d\sigma_{SD}}{dt} \sim F^2(t) \cdot e^{2\alpha' \Delta y t},
\end{equation}
where $F^2(t)$ is given by Eq.~(\ref{eqFF}) and the integration is performed up to $t_{max}=-m^2_p \cdot \frac{\xi^2}{1-\xi}$, with $\xi=e^{-\Delta y}$. The diffractive mass is calculated as $M=\sqrt{s\xi}$. 
The four-momenta of the outgoing proton and the dissociated mass system are calculated using Mandelstam variables for a two-body scattering process, as implemented in {\sc pythia8} for other \texttt{Diffraction:PomFlux} options.

\subsection{Double-diffractive events}
\label{genDD}

Events are generated by first choosing the rapidity-gap width according to Eq.~(\ref{eqDD}) integrated over $t$. Eq.~(\ref{eqDD}) is divergent as $\Delta y \rightarrow 0$. In order to remove the divergence, the integration over $t$ is performed within the limits from $t_{min}=-e^{\Delta y}$ to $t_{max}=-e^{-\Delta y}$. Then, $\Delta y$ is chosen from the distribution:
\begin{equation}
\frac{d\sigma_{DD}}{d\Delta y} \sim e^{\epsilon \Delta y} \cdot \frac{\ln{\frac{ss_0}{M^4_0}}-\Delta y}{2\alpha' \Delta y}\left( e^{-2\alpha'\Delta y e^{-\Delta y}} - e^{-2\alpha'\Delta y e^{\Delta y}}\right) \cdot S,
\end{equation}
and the range of the generation is defined by $\Delta y_{min}=0$ and $\Delta y_{max}=-\ln{M^4_0/(ss_0)}$, where $M^2_0=$\texttt{{\bf\sc mbr}m2Min} and $s_0 = 1~\mbox{GeV}^2$. To further suppress events at low values of $\Delta y$ the term:
\begin{equation}
S=\frac{1}{2}\left[ 1+erf\left(\frac{\Delta y - \texttt{{\bf\sc mbr}dyminDD}}{\texttt{{\bf\sc mbr}dyminSigDD}}\right)\right],
\end{equation}
is used as explained in Sec. \ref{diffXsec}. 

The variable $t$ is chosen according to:
\begin{equation}
\frac{d\sigma_{DD}}{dt} \sim e^{2\alpha' \Delta y t},
\end{equation}
in the range from $t_{min}=-e^{\Delta y}$ to $t_{max}=-e^{-\Delta y}$. 

Then, the center of the rapidity gap, $y_0$, is selected uniformly within the limits: 
\begin{equation}
-\frac{1}{2}\left( \ln{\frac{ss_0}{M^4_0}}-\Delta y\right) < y_0 < \frac{1}{2}\left( \ln{\frac{ss_0}{M^4_0}}-\Delta y\right),
\end{equation}
and the diffractive masses are calculated as:
\begin{equation}
M^2_1=\sqrt{s \cdot e^{-\Delta y - y_0}}, 
\end{equation}
\begin{equation}
M^2_2=\sqrt{s \cdot e^{-\Delta y + y_0}}. 
\end{equation}

The four-momenta of the outgoing dissociated mass systems are calculated using Mandelstam variables for a two-body scattering process, as implemented in {\sc pythia8} for other options of \texttt{Diffraction:PomFlux} .

\subsection{Central-diffractive (DPE) events}
\label{genCD}

Events are generated by first choosing the total rapidity gap width, $\Delta y$, according to Eq.~(\ref{eqCD}), integrated over $t_1$ and $t_2$:
\begin{equation}
\frac{d\sigma_{CD}}{d\Delta y} \sim e^{\epsilon \Delta y} \int^{\Delta y/2-y_0}_{-\Delta y/2+y_0} dy_0 ~ f_{-} \cdot f_{+} \cdot S_1S_2,
\end{equation}
where:
\begin{equation}
f_{\pm} = \left( \frac{a_1}{b_1+\alpha' \Delta y \pm 2\alpha' y_0} + \frac{a_2}{b_2+\alpha' \Delta y \pm 2\alpha' y_0} \right),
\end{equation}
and the integration is performed from $\Delta y_{min}=0$ to $\Delta y_{max}=-\ln{M^2_0/s}$, where $M^2_0=$\texttt{{\bf\sc mbr}m2Min}. For events at low values of $\Delta y$ we suppress individual gaps with the factor: 
\begin{equation}
S=\frac{1}{2}\left[ 1+erf\left(\frac{\Delta y - \texttt{{\bf\sc mbr}dyminCD}/2}{\texttt{{\bf\sc mbr}dyminSigCD}/\sqrt{2}}\right)\right].
\end{equation}

Then, the direction of the centrally-produced hadronic system, $y_c$, is selected uniformly within the region:
\begin{equation}
-\frac{1}{2}\left( \Delta y -  \Delta y_{min}\right) < y_c < \frac{1}{2}\left( \Delta y - \Delta y_{min}\right),
\end{equation}
and rapidity gaps corresponding to each of the two Pomerons are calculated as:
\begin{equation}
 \Delta y_1=\Delta y/2+y_0,
\end{equation}
\begin{equation}
\Delta y_2=\Delta y/2-y_0.
\end{equation}
The four-momentum transfers squared at each proton vertex, $t_1$ and $t_2$, are generated according to:
\begin{equation}
\frac{d\sigma_{CD,i}}{dt} \sim F^2(t_i) \cdot e^{2\alpha' \Delta y_i t_i}, 
\end{equation}
up to $t_{max,i}=-m^2_p \cdot \frac{\xi_i^2}{1-\xi_i}$, where $\xi_i=e^{-\Delta y_i}$ and $i=1,2$. Then, the $p_T$ and $p_z$ of outgoing protons are calculated as $p^2_{T,i}=(1-\xi_i)|t_i|-m^2_p\xi_i^2$ and $|p_{z,i}|=p(1-\xi_i)$, where $p=\sqrt{s/4-m^2_p}$ is the incoming proton momentum.

Finally, the four-momentum of the hadronic system is calculated from the sum of the four-momenta of the Pomerons, each calculated as a difference between the incoming and outgoing proton four-vectors.

\section{{\bf\sc mbr} implementation in {\sc pythia8}}
\label{cards}

The {\bf\sc mbr} generation is activated with the following flag: \texttt{Diffraction:PomFlux = 5}. The simulation works for the $pp$, $p\bar{p}$ and $\bar{p}p$ scattering, and the beam setup is checked in the {\sc pythia8} initialization step.  It is assumed that the user will veto the event generation, after \texttt{Pythia::init() = false} is returned for other beam particles. 

The simulation of the CD (DPE) process is implemented in {\sc pythia8} for the first time. It is activated with the flag \texttt{SoftQCD:centralDiffractive = on}. The corresponding-process number is set to 106, naturally extending the list of soft QCD processes\footnote{103 and 104 for SD with proton dissociation along the positive or negative z-axis, respectively, and 105 for DD.}. In the event record, the outgoing protons' information is written in rows 3 and 4, and the centrally-dissociated hadronic system occupies row 5 (represented by the \texttt{rho\_diff0} pseudo-particle).

Below, we present the default values of parameters, used when the {\bf\sc mbr} simulation is activated:

\begin{itemize}

\item the parameters of the Pomeron trajectory,  $\alpha(t)=1+\epsilon +\alpha't$:\\
\\
\texttt{Diffraction:{\bf\sc mbr}epsilon = 0.104}\\
\texttt{Diffraction:{\bf\sc mbr}alpha = 0.25}

\item the Pomeron-proton coupling, $\beta(0)$ ($\mbox{GeV}^{-1}$), and the total Pomeron-proton cross section, $\sigma_0$ ($\mbox{mb}$), see Sec.~\ref{diffXsec}:\\
\\
\texttt{Diffraction:{\bf\sc mbr}beta0 = 6.566}\\
\texttt{Diffraction:{\bf\sc mbr}sigma0 = 2.82}

\item the lowest mass-squared value of the dissociated system, $M^2_0$, used to evaluate the highest allowed rapidity gap width, $\Delta y_{max}$, see Sec.~\ref{evgen}:\\ 
\\
\texttt{Diffraction:{\bf\sc mbr}m2Min = 1.5}

\item the minimum width of the rapidity gap used in the calculation of $N_{\rm gap}(s)$ (flux renormalization, Sec.~\ref{diffXsec}):\\
\\
\texttt{Diffraction:{\bf\sc mbr}dyminSDflux = 2.3}\\
\texttt{Diffraction:{\bf\sc mbr}dyminDDflux = 2.3}\\
\texttt{Diffraction:{\bf\sc mbr}dyminCDflux = 2.3}

\item the parameters $\Delta y_S$ and $\sigma_S$, used for the cross-section suppression at low $\Delta y$ (non-diffractive region), see Eq.~(\ref{erffun}):\\
\\
\texttt{Diffraction:{\bf\sc mbr}dyminSD = 2.0}\\
\texttt{Diffraction:{\bf\sc mbr}dyminDD = 2.0}\\
\texttt{Diffraction:{\bf\sc mbr}dyminCD = 2.0}\\
\\
\texttt{Diffraction:{\bf\sc mbr}dyminSigSD = 0.5}\\
\texttt{Diffraction:{\bf\sc mbr}dyminSigDD = 0.5}\\
\texttt{Diffraction:{\bf\sc mbr}dyminSigCD = 0.5}

\end{itemize}

In addition, if the option \texttt{Diffraction:PomFlux = 5} is set, no dampening of the diffractive cross sections is performed (no effect of \texttt{SigmaDiffractive:dampen = on}); however, the user can still set his/her own cross sections by hand (\texttt{SigmaTotal:setOwn= on}).

\section{Summary}
We present the {\bf\sc mbr} (Minimum Bias Rockefeller) Monte Carlo simulation, which has been tested using CDF data, and discuss its implementation in the {\sc pythia8} generator. The double-Pomeron-exchange process is included in {\sc pythia8} for the first time. The simulation is designed to apply to all Minimum Bias processes in the LHC energy range. 

\section*{Acknowledgments}
We thank Torbj\"{o}rn Sj\"{o}strand and Stephen Mrenna for many useful discussions and suggestions. We are indebted to Torbj\"{o}rn Sj\"{o}strand for his guidance in implementing the {\bf\sc mbr} code in the {\sc pythia8} framework.



\begin{thebibliography}{99}
  
\bibitem{dinoModel} K. Goulianos, {\em Hadronic diffraction: where do we stand?}, hep-ph/0407035;\\ K. Goulianos, {\em Diffraction in  QCD}, arXiv:hep-ph/020314.
\bibitem{p8ref} T. Sj\"{o}strand, S. Mrenna and P. Skands, JHEP05 (2006) 026, Comput. Phys. Comm. 178 (2008) 852, arXiv:hep-ph/0603175,arXiv:0710.3820.
\bibitem{globalFit} R. J. M. Covolan, K. Goulianos, J. Montanha, Phys. Lett. B {\bf 389}, 176 (1996).
\bibitem{satFroiss} K. Goulianos, {\em Diffraction, Saturation and pp Cross Sections at the LHC}, arXiv:1105.4916.
\bibitem{donnachie} A. Donnachie, P. V. Landshoff, Nucl. Phys. B {\bf 303}, 634 (1988).
\bibitem{cdfSD} K. Goulianos, Phys. Lett. {\bf B358}, 379 (1995).
\bibitem{cdfDD} T. Affolder et al. (CDF Collab.), {\em Double Diffraction Dissociation at the Fermilab Tevatron Collider}, Phys. Rev. Lett. {\bf 87}, 141802 (2001).
\bibitem{cdfCD} D. Acosta et al. (CDF Collab.), {\em Inclusive double Pomeron exchange at the Fermilab Tevatron $\bar{p}p$ collider}, Phys. Rev. Lett. {\bf 93}, 141601 (2004).
\bibitem{kappa} R. J. M. Covolan, K. Goulianos, J. Montanha, Phys. Rev. D {\bf 59}, 114017 (1999).
\bibitem{erfFun} \url{http://en.wikipedia.org/wiki/Error_function}
\bibitem{coherLimit} K. Goulianos, {\em Diffractive interactions of hadrons at high energies}, Phys. Rep. 101, 169 (1983).
  
\end{thebibliography}
\end{document}